\documentstyle[fleqn,a4,12pt]{article}
\newcounter{abc}

\newcounter{Rom}

\newcounter{rom}

\newcounter{nuljed}

\begin{document}

\Large
\begin{center}
   {\bf  A role of antiquantum bits for superdense coding  } \\
              {\bf and quantum computing }  \\
   \vspace*{2ex}
                   { J. Hruby}  \\
   {Group of Cryptology}        \\
   {Union of Czech Mathematicians and Physicists}  \\
   { P.O.B.21 OST, 170 34 PRAHA 7, Czech Republic}   \\
\vspace*{2ex}
\begin{abstract}
  We define an anti-quantum bit state via analogous way as the anti-state
  in Particle Physics.We show the quantum information Feymann diagrams for the teleportations and superdense
coding, which preserve the information flow.

The role of information vacuum and anihilation between error information and
anti-information is mentioned for the stability of quantum
computing.

  \end{abstract}
\end{center}

\vspace*{2ex}

\normalsize

\section{Introduction}

       A quantum bit (qubit) is a quantum system with a two-dimensional
Hilbert space, capable of existing in a superposition of Boolean states
and of being entangled with the states of other qubits [1].

More precisely a qubit is the amount of the information which is contained
in a pure quantum state from the two-dimensional Hilbert space ${\cal H}_2$.

A general superposition state of the qubit is
\begin{equation}
 |\psi\rangle  = {\psi}_0 |0\rangle  + {\psi}_1 |1\rangle  ,
\label{1.1}
\end{equation}
where  ${\psi}_0$  and   ${\psi}_1$ are complex numbers, $|0\rangle$
and $|1\rangle$ are kets representing two Boolean states.
The superposition state has the propensity to be a $0$ or a $1$ and
${|{\psi}_0|}^2 + {|{\psi}_1|}^2 = 1$.

The eq.(1) can be written as
\begin{equation}
 |\psi\rangle  = {\psi}_0\left( \begin{array}{c} 1 \\ 0 \end{array} \right)
 + {\psi}_1 \left( \begin{array}{c} 0 \\ 1 \end{array} \right)  ,
\label{1.2}
\end{equation}

where we labeled $\left( \begin{array}{c} 1 \\ 0 \end{array} \right) $ and
$\left( \begin{array}{c} 0 \\ 1 \end{array} \right)$ two basis states zero
and one.

The Clifford algebra relations of the $2\times2$ Dirac matrices is
\begin{equation}
\left\{{\gamma}^\mu,{\gamma}^\nu \right\} = 2 {\eta}^{\mu\nu} ,
\label{1.3}
\end{equation}
where
\begin{equation}
 {\eta}^{\mu\nu}    =  \left( \begin{array}{cc}
                              -1 & 0 \\
                               0 & 1 \end{array} \right).
\label{1.4}
\end{equation}

We choose the representation
\begin{equation}
 {\gamma}^0 =i{\sigma}^2   =  \left( \begin{array}{cc}
                              0 & 1 \\
                             -1 & 0 \end{array} \right)
\label{1.5}
\end{equation}
and
\begin{equation}
 {\gamma}^1 ={\sigma}^3   =  \left( \begin{array}{cc}
                              1 & 0 \\
                              0 &-1  \end{array} \right) ,
\label{1.6}
\end{equation}
where $\sigma$ are Pauli matrices and ${\gamma}^5={\gamma}^0{\gamma}^1$ .

 The projectors have the form
\begin{equation}
P_0 = \frac{1+{\gamma}^1}{2} =  \left( \begin{array}{cc}
                              1 & 0 \\
                              0 & 0  \end{array} \right)
 \label{1.7}
\end{equation}
and
\begin{equation}
P_1 = \frac{1-{\gamma}^1}{2} =  \left( \begin{array}{cc}
                              0 & 0 \\
                              0 & 1  \end{array} \right)
 \label{1.7}
\end{equation}

These projectors project qubit on the basis states zero
and one:
\begin{equation}
P_0  |\psi\rangle  = {\psi}_0\left( \begin{array}{c} 1 \\ 0 \end{array} \right),
P_1  |\psi\rangle  = {\psi}_1\left( \begin{array}{c} 0 \\ 1 \end{array} \right)
\label{1.9}
\end{equation}
and represent the physical measurements - the transformations of qubits to the
classical bits.

In classical information theory the Shannon entropy is well defined :
\begin{equation}
S_{CL}(\Phi) =-\sum_{\phi} p(\phi) {\log}_2 p(\phi),
\label{1.10}
\end{equation}
where the variable $\Phi$ takes on value $\phi$ with probability  $p(\phi)$
and it is interpreted as the uncertainty about $\Phi$.

The quantum analog is the von Neumann entropy $S_{q}({\rho}_{\Psi})$ of a quantum
state $\Psi$ described by the density operator ${\rho}_{\Psi}$:
\begin{equation}
S_Q(\Psi) = - {Tr}_{\Psi}\left[ {\rho}_{\Psi} {\log}_2{\rho}_{\Psi} \right],
\label{1.11}
\end{equation}
where ${Tr}_{\Psi}$  denotes the trace over the degrees of freedom associated
with $\Psi$.

The von Neumann entropy has the information meaning, characterizing (asymptotically)
the minimum amount of quantum resources required to code an ensemble of quantum
states.

The density operator ${\rho}_{\psi}$ for the qubit state $ |\psi\rangle $ in (1)
is given:

\begin{equation}
\rho = |\psi\rangle \langle\psi| = {|\psi_0|}^2|0\rangle \langle0| +
\psi_0{\psi_1}^* |0\rangle \langle1| + {\psi_0}^*\psi_1 |1\rangle \langle0| +{|\psi_1|}^2|1\rangle \langle1|
\label{1.12}
\end{equation}
and corresponding density matrix is
\begin{equation}
{\rho}_{kl} = \left( \begin{array}{cc}
                              {|\psi_0|}^2 & \psi_0{\psi_1}^*  \\
                           {\psi_0}^*\psi_1 & {|\psi_1|}^2  \end{array} \right)
\label{1.13}
\end{equation}
and $k,l=0,1.$

The von Neumann entropy reduces to a Shannon entropy if ${\rho}_{\Psi}$ is a mixed
state composed of orthogonal quantum states.

In the interesting work [2], N.J.Cerf and C.Adami proposed the conditional
von Neumann entropy as follows:

for the combined system from two states $|i\rangle$  and $|j\rangle$
the von Neumann entropy
\begin{equation}
S_Q(|ji\rangle) = - Tr\left( {\rho}_{|ji\rangle} {\log}_2{\rho}_{|ji\rangle} \right),
\label{1.14}
\end{equation}
\begin{equation}
S_Q(|i\rangle) = - Tr\left( {\rho}_{|i\rangle} {\log}_2{\rho}_{|i\rangle} \right)
\label{1.15}
\end{equation}
and
\begin{equation}
{\rho}_{|j\rangle} = - {Tr}_{|i\rangle}\rho_{|ji\rangle}.
\label{1.16}
\end{equation}

The von Neumann conditional entropy has the form
\begin{equation}
S_Q(|j\rangle | |i\rangle) = - Tr\left( {\rho}_{|ji\rangle} {\log}_2{\rho}_{|j\rangle | |i\rangle} \right).
\label{1.17}
\end{equation}

   The appearence of negative values for the von Neumann conditional entropy
follows from (17), where the  conditional density matrix ${\rho}_{|j\rangle | |i\rangle}$
is based "conditional amplitude operator" [2]:
\begin{equation}
{\rho}_{|j\rangle | |i\rangle} = \exp \left(- \sigma_{|ji\rangle} \right),
\label{1.18}
\end{equation}
where
\begin{equation}
\sigma_{|ji\rangle} = I_{|j\rangle}\ln{\rho}_{|i\rangle} - \ln {\rho}_{|ji\rangle}
\label{1.19}
\end{equation}
and  $I_{|j\rangle}$ being the unit matrix.

Because the information cannot be negative the question arise what precisely
it means.

The quantum entropy of a given quantum state (1) in ${\cal H}_2$ is the difference
between the maximum of the information contained in (1) and information of the information vacuum.

For the qubit
\begin{eqnarray}
S_Q(|\psi\rangle | |v\rangle)=-Tr\left( {\rho}_{|\psi v\rangle} {\log}_2{\rho}_{|\psi\rangle | |v\rangle} \right) =\\
=S_Q(\psi|v)=S_Q(\psi v)- S_Q(v):=S_Q(\psi)=1,
\label{1.20}
\end{eqnarray}
where $|v\rangle$ is the information vacuum state $S_Q(v)=0$.

Now we can start to think about anti-quantum bits as the anti-quantum state in
the dual Hilbert space $\tilde{\cal H}_2$.

\section{Anti-quantum bits}
We shall define anti-quantum bits (antiqubits) by analogy with antiparticles,
as a quantum state from the dual
Hilbert space, capable of existing in a superposition of Boolean states
and of being entangled with the states of other qubits.

This definition agree formally with these previous ones in [2,3],
 where antiqubit is called as a quantum of negative information, which is
 equivalent to a qubit traveling backwards in time.

But here the information of antiqubit is positive only entropy $S_Q(|\bar{\psi\rangle} | |0\rangle)=S_Q(\bar{\psi}):=-1 $,
where $|\bar{\psi\rangle} $ denotes antiqubit state.

For the antiqubit
\begin{eqnarray}
S_Q(|\bar{\psi}\rangle | |v\rangle)=-Tr\left( {\rho}_{|\bar{\psi} v\rangle} {\log}_2{\rho}_{|\bar{\psi}\rangle | |v\rangle} \right)= \\
=S_Q(\bar{\psi}|v)=S_Q(\bar{\psi} v)- S_Q(v):=S_Q(\bar{\psi})=-1.
\label{1.21}
\end{eqnarray}

To obtain the form of $|\bar{\psi\rangle}$ we use the Dirac adjoint and old
ideas about anti-states from Particle Physics.

The antiparticles positrons was defined by Dirac as the energy holes so we
can define the antiqubits as the information holes. As the mass in the Dirac
eq. has the opposite sign here the entropy of qubit and antiqubit has the
opposite sign.

By analogy with the Dirac
adjoint for the two component spinor, we define antiqubit $\bar{\psi} $ as follows:
\begin{equation}
\bar{\psi}=\psi^+ {\gamma}^0 =\left({\psi_0}^*,{\psi_1}^* \right) \left( \begin{array}{cc}
                              0 & 1 \\
                             -1 & 0 \end{array} \right)
= \left({-\psi_1}^*,{\psi_0}^* \right) = {\psi_0}^* \langle 1| - {\psi_1}^* \langle 0| .
\label{2.1}
\end{equation}
The corresponding density matrix for the antiqubit $\bar{\psi}$ is
\begin{equation}
{{\rho}_{kl}}^` = \left( \begin{array}{cc}
                              {|\psi_0|}^2 & -\psi_0{\psi_1}^*  \\
                              -{\psi_0}^*\psi_1 & {|\psi_1|}^2  \end{array} \right)
\end{equation}
and $k,l=0,1.$

The information vacuum is $|v\rangle \in {\cal H}_2\otimes\tilde{\cal H}_2$.

The two members of a spatially separated Einstein-Podolsky-Rosen (EPR) pair are
maximally entangled qubits so called ebits [2,4].
EPR pair $e\bar{e}$ is created from the information vacuum, which is the pure
quantum state without information $S_Q(\psi_e\psi_{\bar{e}})=S_Q(\psi_e \bar{\psi_e})=S_Q(e\bar{e})=S_Q(|v\rangle)=0$.

As the EPR pairs $e\bar{e}$ after creation from the information
vacuum contain no readable information they represents the virtual information
elementary objects.
The conditional quantum entropy between $e$ and $\bar{e}$ is
 $S_Q(e|\bar{e})=S_Q(e\bar{e})-S_Q(\bar{e})=0-(-1)=1$,

$S_Q(\bar{e}|e)=S_Q(\bar{e}e)-S_Q(e)=0-1=-1$.

Quantum information processes
can be described by quantum information diagrams similar to Feymann diagrams in
Particle Physics, where elementary objects interact.
The information flow is conserved in each vertex of these diagrams.
The law of the conservation of quantum information has no analog in classical
information

The elementary objects of the quantum information dynamics (QID) can be
classical bits $c$, qubits $q$, antiqubits $\bar{q}$, ebits $e$ and antiebits
 $\bar{e}$. But there could be some more elementary virtual quantum information
 objects "the information quarks".

Moreover these virtual elementary objects of the  QID can be combined into the
more complicated information objects in analogy with quarks in the bag models.
There is shown that quarks play no role for the energy of the bag [5] and so virtual
quantum information entangled quarks (equarks)
play no role for the information of the information bag.

\section{Feymann quantum information diagrams}

We show QID for the information processes teleportation  and  superdense
coding, which are connected via time reversal operation.

In quantum teleportation [see Fig.1] a qubit $\psi_q$ is transported with
perfect fidelity between two vertices $M$ and $U$ through the transmission
of two classical $2c$ bits and shared EPR pair $e\bar{e}$:

\setlength{\unitlength}{1mm}
\begin{picture}(40,50)
\put(12,20){$\psi_q$}
\put(33,29){$\psi_e$}
\put(33,9){$2c$}
\put(33,12){$\succ$}
\put(33,26){$\prec$}
\put(20,19){$\succ$}
\put(48,19){$\succ$}
\put(35,20){\circle{30}}
\put(18,20){\line(1,0){10}}
\put(28,20){\circle*{1}}
\put(42,20){\circle*{1}}
\put(52,20){\circle*{1}}
\put(18,20){\circle*{1}}
\put(42,20){\line(1,0){10}}
\put(30,19){$M$}
\put(38,19){$U$}
\put(55,20){$\psi_q$}
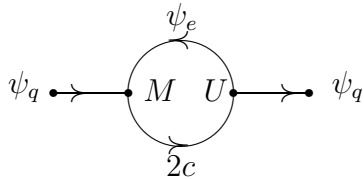
\put(1,1){Fig.1 Feymann quantum information diagram for teleportation.}
\end{picture}

Here $M$ (the measurement) and $U$ (the unitary
transformation) mean the two processes in QID and their description
is written elsewhere [6].
In vertex M qubit interacts with ebit and it gives four orthogonal maximally
entangled $|eq\rangle$ states, which are transported via 2 cbits.

In vertex U is the reconstruction of ${\psi}_q$ the  qubit from 2 c bits
via applying to antiebit one of four possible unitary transforms in the
one state Hilbert space ${\cal H}_2$.

The information flow is conserved in each vertex of the diagram on Fig.1.
For the vertex M the entropy conservation rule is
\begin{equation}
S_Q(q)+S_Q(e) = 1 + 1 = 2
\label{3.1}
\end{equation}
since qubit and ebit are initially independent.
At vertex U we have
\begin{equation}
S_Q(q)=S_Q(2c)+S_Q(\bar{e})=2-1=1=S_Q(qe\bar{e})=S_Q(qe) + S_Q(\bar{e}|qe).
\label{3.2}
\end{equation}
From eq.(2) we can see that outgoing $e$ from vertex U  is equivalent
to the incoming $\bar{e}$ to U.

In superdense coding [see Fig.2] we transport 2 cbits via 1 qubit.
The two cbits can be packed into one qubit via applying the unitary
transform what is equivalent to the interaction $2c$ with $\bar{e}$.
At vertex M the interaction between qubit and ebit gives recovering
of the 2 cbits.

\setlength{\unitlength}{1mm}
\begin{picture}(50,50)
\put(12,20){$2c$}
\put(33,29){$\bar{\psi_e}$}
\put(33,9){$\psi_q$}
\put(33,12){$\succ$}
\put(33,26){$\prec$}
\put(20,19){$\succ$}
\put(48,19){$\succ$}
\put(35,20){\circle{30}}
\put(18,20){\line(1,0){10}}
\put(28,20){\circle*{1}}
\put(42,20){\circle*{1}}
\put(52,20){\circle*{1}}
\put(18,20){\circle*{1}}
\put(42,20){\line(1,0){10}}
\put(30,19){$U$}
\put(37,19){$M$}
\put(55,20){$2c$}
\put(1,1){Fig.2 Feymann quantum information diagram for superdense coding.}
\end{picture}

The diagramm in Fig.2 is equivalent to the Feymann  quantum information diagram where ebit
$\psi_e$ is sent backwards in time ( it is equivalent to antiebit $\bar{\psi_e}$):

\begin{picture}(50,50)
\put(12,20){$2c$}
\put(33,29){$\psi_e$}
\put(33,9){$\psi_q$}
\put(33,12){$\succ$}
\put(33,26){$\succ$}
\put(20,19){$\succ$}
\put(48,19){$\succ$}
\put(35,20){\circle{30}}
\put(18,20){\line(1,0){10}}
\put(28,20){\circle*{1}}
\put(42,20){\circle*{1}}
\put(52,20){\circle*{1}}
\put(18,20){\circle*{1}}
\put(42,20){\line(1,0){10}}
\put(30,19){$U$}
\put(37,19){$M$}
\put(55,20){$2c$}
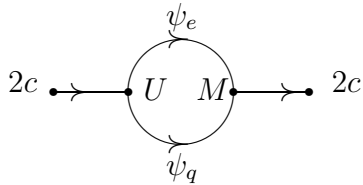
\put(1,1){Fig.3 Feymann quantum information diagram for superdense coding with $\psi_e$.}
\end{picture}

In this case at vertex U we have
\begin{equation}
S_Q(2c) + S_Q(\bar{e})= 2-1=1=S_Q(q|e)=S_Q(2c\bar{e}|e)=S_Q(2c)+S_Q(\bar{e}|e),
\label{3.3}
\end{equation}
since 2c and $\bar{e}$ are initially independent.
At M is
\begin{equation}
S_Q(2c)=S_Q(qe)=S_Q(q|e) +S_Q(e) = 1+1=2.
\label{3.3}
\end{equation}

The quantum information conservation law can have important consequences in
cosmology.

\section{Quantum error correction using antiqubits}

The main problem in quantum computing are error processes, such as decoherence
and spontaneous emission, on the state of a quantum memory register. For
example if a quantum memory register is described by an isolated qubit (1)
with the density matrix (13) at initial moment, the density matrix changes
over time:
 \begin{equation}
{\rho}_{kl} = \left( \begin{array}{cc}
                              {|\psi_0|}^2 &e^{-\frac{t}{\tau}}\psi_0{\psi_1}^*  \\
                          e^{-\frac{t}{\tau}} {\psi_0}^*\psi_1 & {|\psi_1|}^2  \end{array} \right)
\label{1.13}
\end{equation}
and $k,l=0,1.$

Here $\tau$, called the "decoherence time", sets the characteristic time-scale
of the decoherence process.

 For the antiqubit (21) we obtain

\begin{equation}
{{\rho}_{kl}}^` = \left( \begin{array}{cc}
                              {|\psi_0|}^2 & -e^{-\frac{t}{\tau}}\psi_0{\psi_1}^*  \\
                              -e^{-\frac{t}{\tau}}{\psi_0}^*\psi_1 & {|\psi_1|}^2  \end{array} \right)
\end{equation}
and $k,l=0,1.$

Quantum error-correction strategy can be based on the idea of mirror computers,
which are working in paraell regime with qubits and antiqubits.

It is known that quantum error-correcting code [7] works on the idea
that several physical qubits are used to encode one logical qubit.The actual
value of the logical qubit is stored in the correlations between the several
classical physical qubits so that even if there is disruption to a few
qubits in the codewords, there is still sufficient information in the
correlation when error occurr and then correct the qubit.

In mirror computation we can use the correlation between ebits and antiebits
to anihilate the error information.

We know that any error in a single bit can be described by the action
of linear combination of Pauli matrices:

\begin{equation}
{\sigma}_1 \left( \begin{array}{c} {\psi}_0 \\ {\psi}_1 \end{array} \right)=\left( \begin{array}{cc}
                               0 & 1 \\
                               1 & 0 \end{array} \right)\left( \begin{array}{c} {\psi}_0 \\ {\psi}_1 \end{array} \right)=\left( \begin{array}{c} {\psi}_1 \\ {\psi}_0 \end{array} \right),
\label{3.3}
\end{equation}
i.e. bit flip error,

\begin{equation}
{\sigma}_2 \left( \begin{array}{c} {\psi}_0 \\ {\psi}_1 \end{array} \right)=\left( \begin{array}{cc}
                               0 & -i \\
                               i& 0 \end{array} \right)\left( \begin{array}{c} {\psi}_0 \\ {\psi}_1 \end{array} \right)=i\left( \begin{array}{c} {\psi}_1 \\ {\psi}_0 \end{array} \right),
\label{3.3}
\end{equation}
i.e. phase shift plus bit flip error,

\begin{equation}
{\sigma}_3 \left( \begin{array}{c} {\psi}_0 \\ {\psi}_1 \end{array} \right)=\left( \begin{array}{cc}
                               1&0 \\
                               0&-1 \end{array} \right)\left( \begin{array}{c} {\psi}_0 \\ {\psi}_1 \end{array} \right)=\left( \begin{array}{c} {\psi}_0 \\-{\psi}_1 \end{array} \right),
\label{3.3}
\end{equation}
i.e. phase shift error.

For antiqubits we obtain the same in the dual space.

So the error correction codes using the antiqubits are based on double symmetrization.
The first one is going via correlation between physical qubits and  the second
one between logical ebits and antiebits via the anihilation
of the error information $S_Q(e\bar{e})=0$. The work on this realization is in
progress.

\section{Conclusions}
We give in this paper the definition of the antiqubit via analogy with
antistates in Particle Physics.

We mentioned some new application of the antiqubits and the quantum information
conservation law.

Some speculations with the connection of the qubits and antiqubits with
non-commutative geometry, quantum space-time, superstrings and membranes
theory can be done, to obtain information connected with the geometry and
not with the particles.

   This work was supported by  Grant RN 19982003013.

\end{document}